\documentclass[twocolumn,prl,superscriptaddress,nofootinbib]{revtex4}
\pdfoutput=1

\usepackage[english]{babel}
\usepackage{amsmath}
\usepackage{graphicx}
\usepackage{amsbsy, amstext}
\usepackage{hyperref}
\usepackage{amsmath, amsthm, amssymb, upref, mathrsfs, amsfonts}
\usepackage{epstopdf}
\usepackage{color}

\newcommand{\pe}{\phi_{\mathrm{eff}}}
\newcommand{\Mpl}{M_{P}}
\newcommand{\U}{\mathcal{U}_{mod}}
\newcommand{\be}{\begin{eqnarray}}

\newcommand{\V}{\mathcal{V}}

\newcommand{\ee}{\end{eqnarray}}
\newcommand{\bdm}{\begin{displaymath}}
\newcommand{\edm}{\end{displaymath}}
\newcommand{\cu}{r}
\newcommand{\cd}{{\theta}}
\newcommand{\fu}{f_r}
\newcommand{\fd}{f_{\theta}}
\newcommand{\tcd}{\tilde{\theta}}
\newcommand{\tcu}{\tilde{r}}

\newcommand{\al}{\alpha'}
\newcommand{\cuin}{r_{\mathrm{in}}}

\begin{document}

\title{Dante's Inferno}%

\author{Marcus Berg}
\affiliation{Oskar Klein Center for Cosmoparticle Physics 
and Department of Physics, 
Stockholm University, Albanova University Center, SE-106 91 Stockholm, Sweden}

\author{Enrico Pajer}
\affiliation{Institute for High Energy Phenomenology,
Newman Laboratory of Elementary Particle Physics,
Cornell University, Ithaca, NY 14853, USA}

\author{Stefan Sj\"ors${}^{1,2}$}

\begin{abstract}
We present a simple two-field model of inflation and show how to embed it in string theory as a straightforward generalization of axion monodromy models. Phenomenologically, the predictions are equivalent to those of chaotic inflation, and in particular include observably large tensor modes. The whole high-scale large-field inflationary dynamics takes place within a region of field space that is parametrically subplanckian in diameter, hence improving our ability to control quantum corrections and achieve slow-roll inflation. 
\end{abstract}

\maketitle

%The study of the early universe provides a way to test ideas about the laws of nature at extraordinary high energies. 
In the observationally successful framework of inflationary cosmology, we can study detectable consequences of physics at very high energies. The energy 
scale at which inflation took place is a free parameter that is still very poorly constrained, and can range from the $\mathrm{GUT}$ scale to the $\mathrm{TeV}$ scale (or even less). The high-scale end of this interval is the most exciting: first, it is the closest to the Planck scale\footnote{We use the reduced Planck mass defined by $\Mpl^2\equiv(8\pi G_N)^{-1}$.} $\Mpl$, where quantum gravity should become important; second, the scale of inflation is determined by the amplitude of primordial tensor modes, which will be detectable in the CMB only if this scale is close to the GUT scale.

An interesting perspective on high-scale inflation is given by the Lyth bound \cite{Lyth:1996im}. This says that detectably-large tensor modes, which are equivalent to high scale inflation\footnote{We assume perturbations are generated by the inflaton.}, require \textit{superplanckian} variation of the inflaton field. This increases the UV-sensitivity of inflation, e.g.~in the sense that an infinite sequence of Planck-suppressed higher dimension operators become crucial for assessing the success of the model. This suggests that high-scale large-field models provide a framework particularly well suited to test candidate UV completions of quantum field theory plus general relativity, such as string theory.

In this work we study a model of inflation, which we call Dante's Inferno, where high-scale large-field inflationary dynamics takes place within a region of field space which is parametrically \textit{subplanckian} in diameter. This provides a new perspective on the Lyth bound and its implications. We organize our presentation as follows. First, we describe the effective field theory implementation of this model, which contains a mechanism to alleviate any $\eta$-problem that might be present. Then we show how Dante's Inferno can be embedded in string theory,
as a straightforward generalization of axion monodromy models. The multi-field dynamics alleviates two of the leading backreaction constraints present in the single-field case.

%%%%%%%%%%%%%%%%%%%%%%%%%%%%%%%%%%%%%%%%%%%%%%%%%%%%%%%%%

\section{The effective field theory model}
\begin{figure}
\begin{center}
\includegraphics[width=60mm]{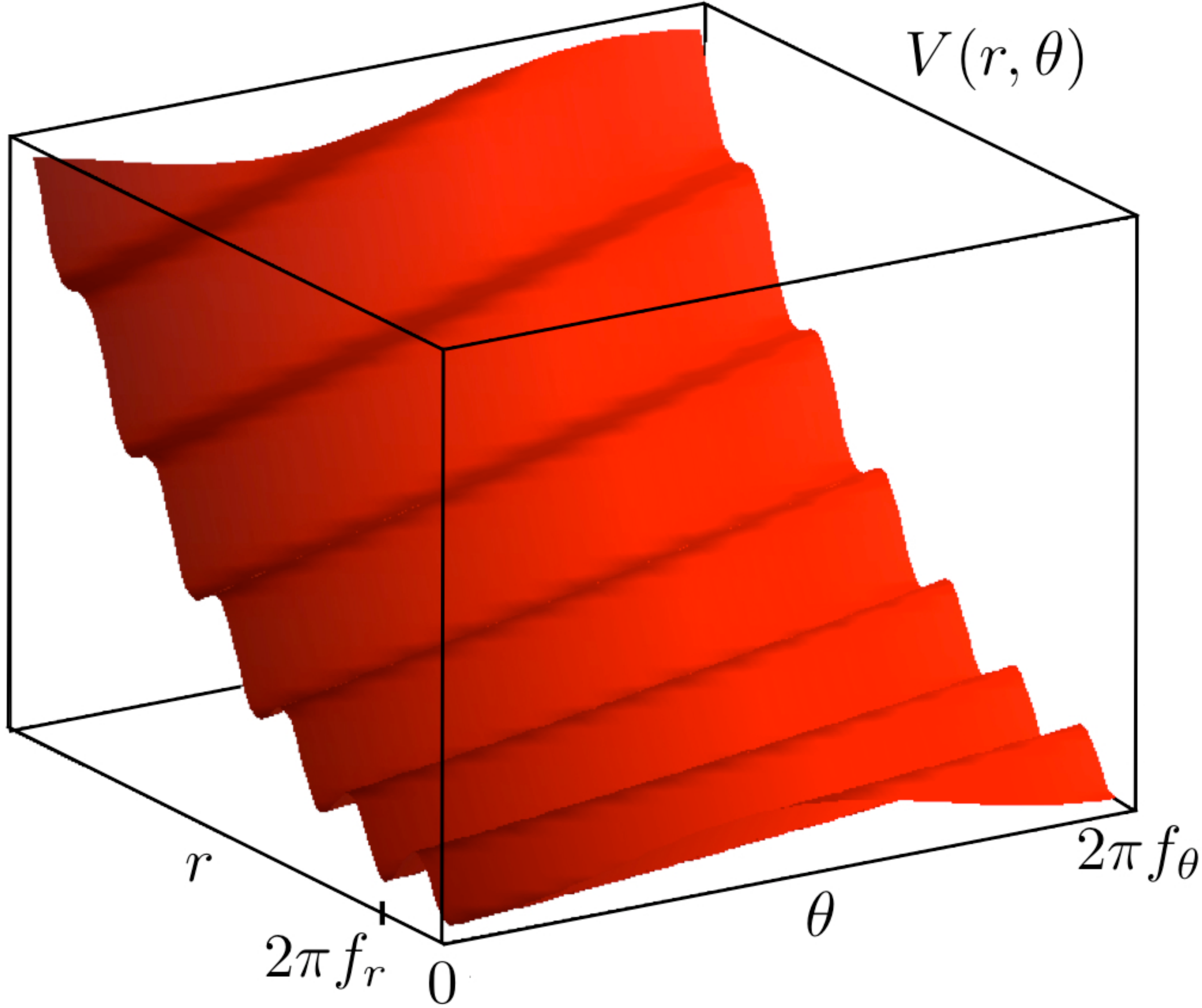}
\caption{The potential in \eqref{V}  (for $W(\cu)=\frac12 m^2 \cu^2$) in Cartesian coordinates, which faithfully represent the metric on field space.\label{fig2}}
\end{center}
\end{figure} 

Our effective model consists of two axions $\cu$ and $\cd$ whose decay constants $\fu$ and $\fd$ obey $\fu<\fd$. A linear combination of  $\cu$ and $\cd$ receives a periodic potential from some non-perturbative effect, which breaks their continuous shift symmetry down to a discrete one. In addition, we explicitly break the shift symmetry for $\cu$ by introducing a term $W(\cu)$ in the potential, where $W(\cu)$ is an a priori arbitrary
regular function. The resulting Lagrangian for canonically normalized fields is
\be
{\mathcal L} &=& \frac12 (\partial \cu)^2 +\frac12 (\partial \cd)^2-V(\cu,\cd)\,,\label{L}\\
 V(\cu,\cd) &=& W(\cu) + \Lambda^4\left[1- \cos\left( \frac{\cu}{\fu} - \frac{\cd}{\fd} \right)\right]\,,\label{V}
\ee
where $\Lambda$ is a non-perturbatively generated scale and $\cu$ and $\cd$ have dimensions of mass. Notice that in the cosine we have chosen a particularly simple linear combination. At the level of the effective action this can be done without loss of generality, since other values of the
coefficients could be reabsorbed in the definition of $\fu$ and $\fd$. On the other hand, in the string theory construction, $\fu$ and $\fd$ are determined by the geometry of the compactification. We will come back to this point in the next section. We define $W(\cu)$ such that the vacuum energy density vanishes at the minimum.
 
We have performed our computations with
$W(\cu)$ 
an arbitrary monomial in $\cu$ and we will give explicit formulae for this 
 at the end of this section.
In the axion monodromy string models (about which more in the next section), so far only 
the linear case $W(\cu)=\mu ^3 \cu$ has been studied,
where $\mu$ is a constant. 
For the purpose of exposition, in the following we will discuss 
details for the quadratic case $W(\cu)= \frac{1}{ 2}m^2 \cu^2$ instead,
for two reasons. First, it captures a few minor additional complications which are absent in the linear case only. Second, it is the archetypal potential with  an $\eta$ problem. 
% We note already here that one can think of the quadratic case as a Taylor expansion of a generic $W(\cu)$,
%since as we will see, the whole  inflationary dynamics relevant for observations takes place within a relatively narrow range $\Delta\cu\ll\Mpl$. 

We will soon find out that the whole observable inflationary dynamics involves a range $d_{\cu}$ that can be made parametrically small. If $d_{\cu}$ were arbitrarily small, we could always expand any $W(\cu)$ around a minimum, hence obtaining a quadratic potential. In concrete realizations, e.g.~the string theory model presented in the next section, the various parameters are constrained and it is a model-dependent question whether the quadratic approximation is accurate. In any event, the validity of our mechanism does not rely on  $W(\cu)$ being quadratic.

We would also like to stress that the field space metric is simply $d\cu^2+d\cd^2$, as can be seen from \eqref{L}, and not the induced flat space metric in polar coordinates.

\begin{figure}[h]
\begin{center}
\includegraphics[width=55mm]{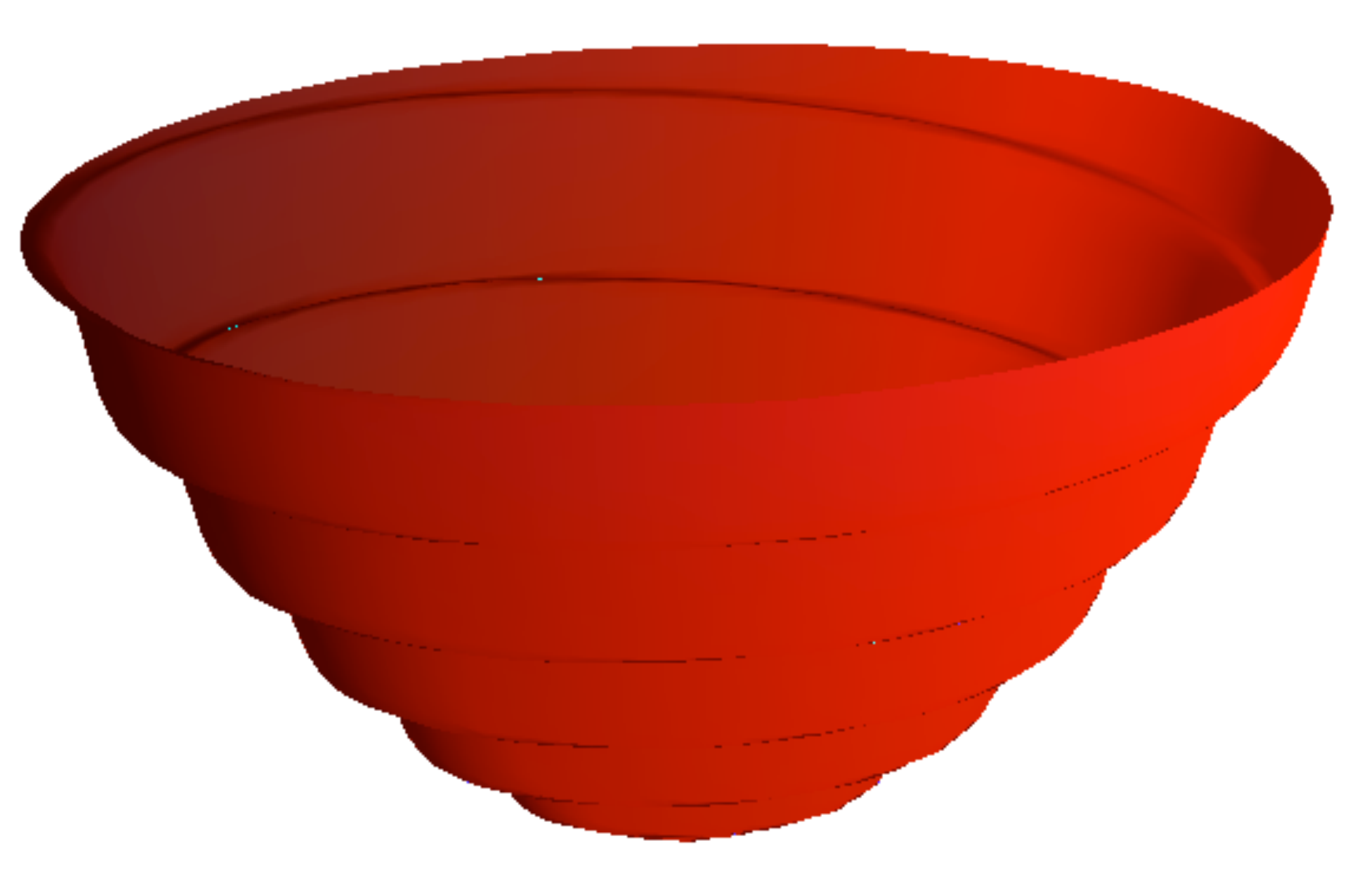}
\includegraphics[width=55mm]{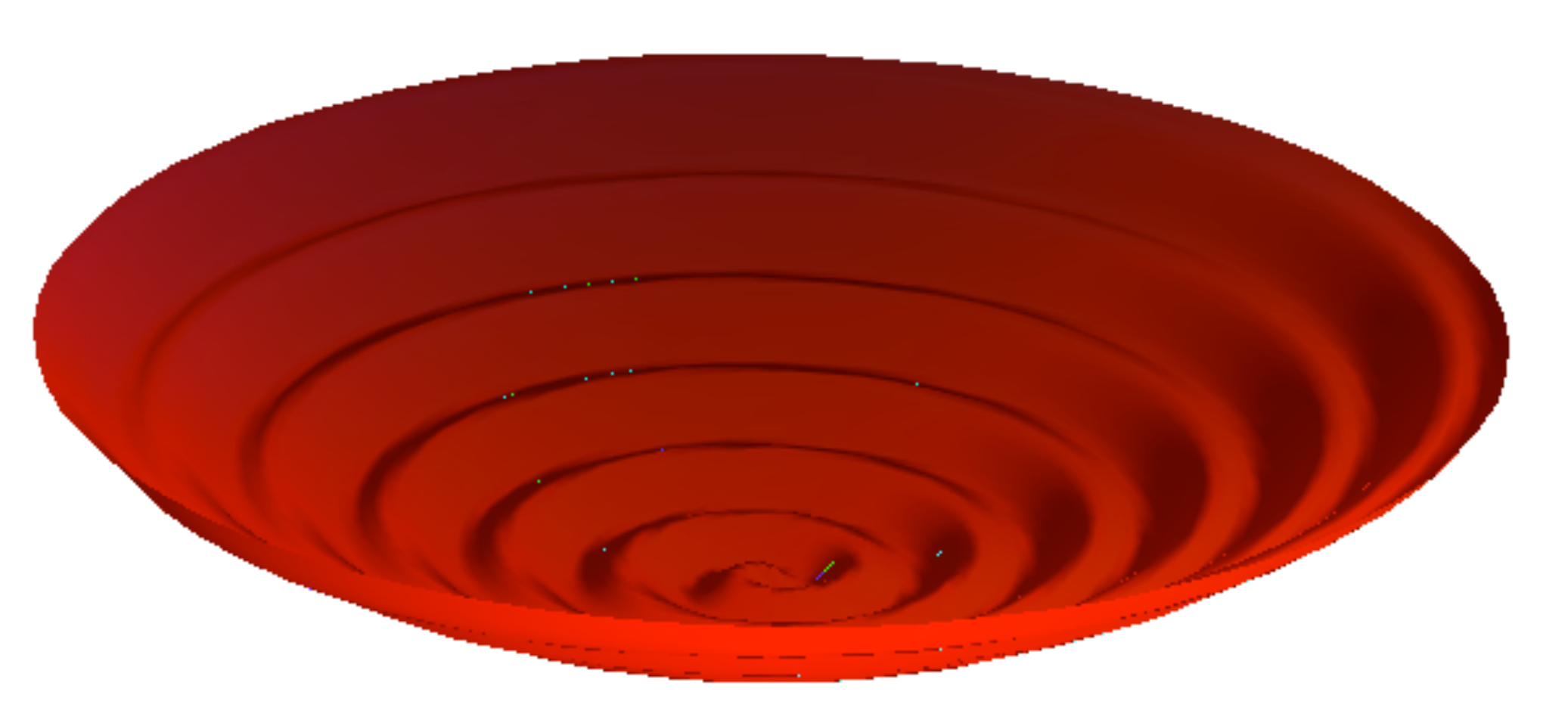}
\caption{The potential in \eqref{V} for $W(\cu)=\frac12 m^2 \cu^2$ in cylindrical coordinates, which \textit{do not} faithfully reproduce the field-space metric as given in \eqref{L}, contrary to Cartesian coordinates in figure \ref{fig2}. On the other hand, in polar coordinates, the periodicity in $\cd$ is apparent and the similarity with Inferno as described by Dante \cite{Dante} becomes evident. \label{fig1}}
\label{pot2}
\end{center}
\end{figure}

We plot the potential as in figure \ref{fig2}. It is clear that
the distance traversed along each ``trench" is (almost) the same at different heights. 
However, the periodicity in $\theta$ is not manifest. 
We therefore also show 
figure \ref{fig1}, which is the justification for the title of this work: the potential \eqref{V} in polar coordinates looks like a spiral staircase, where the periodicity $\cd\rightarrow\cd+2\pi \fd$ is manifest. 
However, in figure \ref{fig1}, the induced field space metric is not faithfully represented, unlike in fig.\ \ref{fig2}
 (it is $dr^2 + d\theta^2$, as stated above). 
 Indeed, 
in fig.\ \ref{fig1}, one can be misled to believe that the spiral trenches at the bottom are much smaller 
per revolution than those further up, but
as we saw in fig.\  \ref{fig2}  they are actually almost the same length at different heights.

The inflationary dynamics can be intuitively read off from either figure. The system starts somewhere up in the $W(\cu)$ potential, quickly rolls in the $\cu$ direction to the closest trench\footnote{The initial condition problem here is no worse than in standard chaotic inflation. If the inflaton has some downward initial speed, then it needs to start at some larger $\cu$ to achieve enough e-folds of inflation.} and from there it slowly spirals along the trench mostly in the $\cd$ direction. This classical two-field slow-roll motion can  be captured by an effective single-field potential that we are now going to derive.

First it is convenient to rotate our coordinates by
\be
\left(
 \begin{array}{c} 
  \tcd \\ 
  \tcu 
 \end{array}
 \right)=
  \left(
 \begin{array}{cc} 
   \cos \xi & \sin \xi  \\  
   -\sin \xi & \cos \xi
 \end{array}
 \right)
 \left(
 \begin{array}{c} 
   \cd \\  \cu
 \end{array}
 \right)\,,
\ee
with
\be
 \sin\xi = \frac{f_r}{\sqrt{f_r^2 + f_{\theta}^2}}, \quad \cos\xi = \frac{ f_{\theta}}{\sqrt{f_r^2 + f_{\theta}^2}},
\ee
after which the potential becomes
\be \label{V2}
V=\frac 12 m^2 \left(\tcu\cos \xi +\tcd \sin \xi \right)^2+\Lambda^4\left[1- \cos \frac{\tcu}{f} \right]\,,
\ee
where $f=\fu\fd/(\fu^2+\fd^2)^{1/2}$. As initial condition, we assume the system starts high up in the quadratic potential, but still at subplanckian values, i.e. $f\ll \cuin < \Mpl$. An interesting regime to consider is then
\begin{enumerate}
   \item[A.] $\fu\ll\fd \ll \Mpl$, \label{1}
   \item[B.] $\Lambda^4\gg f m^2 \cuin$. \label{2}
\end{enumerate}
It will turn out that useful values of $\Lambda$ range from $10^{-3}M_P$ to $10^{-1} M_P$.
Typical values for $f_r$ and $f_{\theta}$ to keep in mind are $10^{-3}M_P$ and
$10^{-1}M_P$, respectively.
The advantage of considering subplanckian $\fu$ and $\fd$ is that we will eventually embed this model in
controlled string models, where superplanckian axion decay constants seem elusive \cite{Banks:2003sx}. We will assume for the rest of this section that these two conditions hold. Notice that condition A implies
\be 
\cos \xi\simeq1\,,\quad \sin \xi\simeq \frac{\fu}{\fd}\,,\quad f\simeq \fu\,,
\ee 
to leading order in $\fu/\fd$. 

\begin{figure}
\begin{center}
\includegraphics[width=85mm]{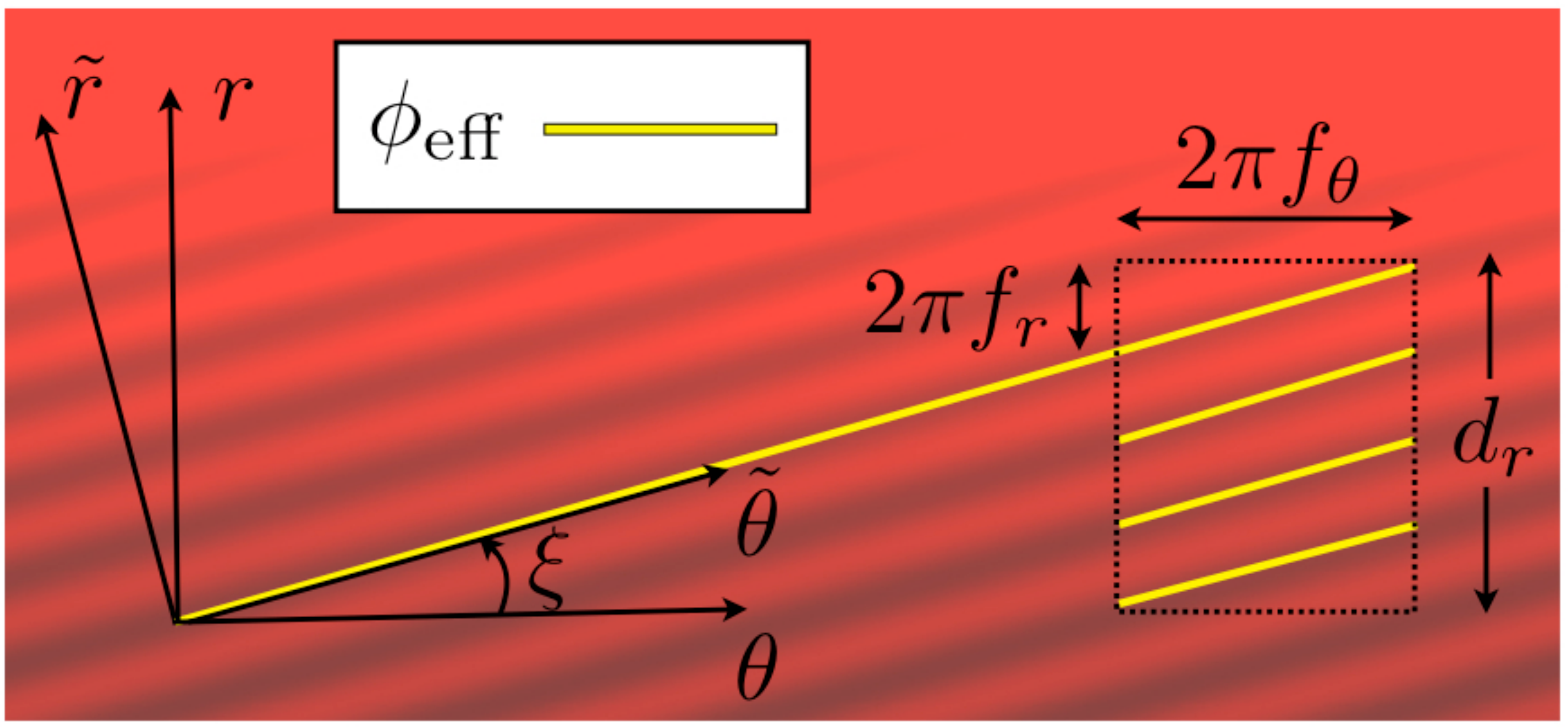}
\caption{The field-space contour plot of the potential. Brighter red shading corresponds to higher energy and vice versa. Both rotated and un-rotated coordinates are shown together with $\pe$ which is always tangent to the inflationary trajectory. Upon the use of the $\cd$ shift-symmetry, the whole inflationary trajectory can be contained into a region of subplanckian diameter, indicated by the dotted box.}
\label{coord}
\end{center}
\end{figure}

When the system rolls along the bottom of a trench, the excitations in the $\tcu$ directions are very massive, i.e.~ $\partial_{\tcu}^2 V>H$, and $\tcu$ can be integrated out. This adiabatic approximation boils down to assuming that, as the system evolves mostly along the trench in the $\tcd$ direction, the new $\tcd$-dependent minimum in the $\tcu$ direction is reached very quickly. So we need to solve $\partial_{\tcu}V=0$, i.e.~
\be \label{ad}
\frac{\Lambda^4}{f}\sin \frac{\tcu}{f}+m^2\cos \xi \left(\tcu \cos \xi +\tcd \sin \xi \right)=0\,,
\ee
or equivalently
\be 
\sin \frac{\tcu}{f}=-\frac{m^2\cos \xi f r}{\Lambda^4}\,.
\ee
Conditions A and B then imply $\sin(\tcu/ f)\ll 1$ at the initial and all subsequent times. Modulo a shift in $\cd$, which is a symmetry of the Lagrangian and in tilded coordinates corresponds to $\tcu \rightarrow \tcu-2\pi \fd \sin\xi$ and $\tcd\rightarrow\tcd+2\pi \fd\cos\xi$, one finds $\tcu\ll f$. Intuitively, this means that $\tcu$ stays at the bottom of the trench throughout the infernal dynamics (see figure \ref{coord}). Hence we linearize \eqref{ad} and find
\be \label{r}
\tcu\simeq -\tcd \frac{m^2 \cos \xi \sin \xi f^2}{m^2 \cos ^2 \xi f^2+\Lambda^4} \simeq - \tcd\frac{\fu}{\fd}\frac{m^2 \fu^2}{\Lambda^4}\,,
\ee
where again we made use of conditions A and B. Using the adiabatic approximation, we can describe the dynamics in the two-field potential \eqref{V} using a single-field potential. 
Renaming $\tcd=\pe$, the solution \eqref{r}  reduces \eqref{V2} to
\be 
V_{\mathrm{eff}}(\pe)=
\frac12 \, m_{\rm eff}^2\,   \pe^2 \; , \quad
m_{\rm eff} \equiv
m\,   \frac{f_r}{f_{\theta} } 
\ee
and the kinetic term is still canonical at leading order in $\fu/\fd$. 

To summarize, we have shown that the Dante's Inferno two-field potential leads to dynamics effectively equivalent to that of the archetypal model of single-field chaotic inflation, $V = \frac{1}{ 2}m^2 \phi^2$, for which the scalar-to-tensor ratio is 0.14 and the spectral tilt is $n_s\simeq 0.96$. However, there are two remarkable differences in the parameters.

First, the effective mass $m_{\rm eff}$ is suppressed by the small factor $\fu/\fd$ with respect to $m$. This  parametrically improves the ability to avoid a potential $\eta$ problem. As a figure of merit, if $m\sim H$ as is typical in supergravity and superstring models, then a modest hierarchy $\fu/\fd\sim \mathcal{O}(10^{-1})$ is sufficient to ensure the validity of the slow-roll conditions.

Second, the interpretation
of the Lyth bound is not the same 
in the original variables $r,\theta$ as in $\phi_{\rm eff}$. We will devote the next subsection to this point. 

Kim, Nilles and Peloso \cite{Kim:2004rp} 
proposed an interesting two-field model
(based on \cite{Freese:1990rb}) that also has the two properties of the previous paragraphs. A drawback of their model is that it relies on a precise cancellation between two parameters (their $g_1$ and $g_2$). As a figure of merit, starting with axion decay constants of order $g_1,g_2\sim10^{-1}\Mpl$, fine tuning on the order of $1\%$ to $0.1 \%$ is needed for their model to work. In Dante's Inferno, no fine tuning is required to make the potential suitable for 60 e-folds of slow-roll inflation. In fact, $\fu$ and $\fd$ can be changed by order-one factors without affecting the main result. Perhaps more importantly, our setup is derived from a concrete string theory construction (in the case $W(r)=\mu^3 r$) that we will describe in the next section.

Finally, we report similar results for  a generic monomial $W(\cu)=\lambda_p \cu^p/p!\,$. In the regime analogous to that given by the conditions A and B above, one finds that the effective single-field potential is
\be 
V_{\mathrm{eff}}= \frac{\lambda_p^{\rm eff}}{p!}\pe^p\; , \quad \label{p}
\lambda_p^{\rm eff}= \left(\frac{\fu}{\fd}\right)^{\! p}\! \lambda_p \; .
\ee
From the string-theory perspective that we will soon adopt, an interesting special case of \eqref{p} is $p=1$, i.e.~$W(\cu)=\lambda_1 \cu$, where $\lambda_1$ is a constant of dimension (mass)$^3$,
called $\mu^3$ in the string models. In this case $\tcu$ can be integrated out exactly, without the need to expand \eqref{ad}. In fact, $\tcu$ is exactly constant throughout the slow-roll inflationary dynamics.
For generic $p$, \eqref{p} shows that even if one starts with a coupling $\lambda_p$  dictated by naturalness, plus the assumption of a $M_P$ (or lower) cutoff, a hierarchical choice of $\fu$ and $\fd$ can still ensure slow-roll inflation. Intuitively, the inflationary direction $\phi_{\rm eff}$ is a mixture of the potentially steep $\cu$ direction and the nearly-flat $\cd$ direction, with a small mixing angle given by $\sin \xi\sim\fu/\fd$. 

\subsection{Field range and Lyth bound}

The Lyth bound \cite{Lyth:1996im} says that observably large tensor modes, equivalent to high-scale inflation, require that the length of the inflationary trajectory is superplanckian. The length is measured by $\Delta \phi_{\rm eff} = \int d\pe$, where $\pe$ is defined as the canonically normalized field always tangent to the inflationary trajectory. In fact, this can be verified in our model, where we found a tensor-to-scalar ratio of 0.14 and $\Delta \pe\sim 15 \Mpl$. The relevance of the Lyth bound lies in its implications for the robustness of the potential against possible corrections. In order to discuss this, let us introduce some more precise language. We call \textit{fundamental fields} those that naturally appear in the Lagrangian as derived from a fundamental UV theory, such as string theory, which have a precise meaning in this UV theory. As we will see, both $\{ \cu,\cd\} $ and $\{ \tcu,\tcd\} $ are fundamental fields. Symmetries, like the $\cd$ shift symmetry in our case, are most explicit in terms of these fields. We call \textit{effective fields} the result of those convenient field redefinitions that are useful to describe particular phenomena. They may not have a precise interpretation in the UV theory and may obscure its underlying symmetries. For example, $\pe$ is convenient to describe the inflationary dynamics, but any notion of a shift symmetry $\theta \rightarrow \theta +$constant is lost in this formulation, as in \eqref{p}.

The Lyth bound is usually supplemented by some prejudices about possible corrections. Without further specification, one generically expects that a large field range opens up the possibility for large corrections, which in turns makes it harder to ensure flatness of the potential, for example. It is appropriate to discuss corrections in terms of fundamental fields, because they have a clear interpretation in the UV theory from which these corrections can come. Also, in terms of these fields, symmetries can be most effectively used to understand and control such corrections,
as usual in effective field theory. In Dante's Inferno, for example, the periodicity in $\cd$ can be used to argue that no polynomial corrections can spoil the flatness in that direction. On the other hand, the Lyth bound is a statement about the range of effective fields. This dichotomy is latent in single-field models but becomes particularly sharp in multi-field models such as Dante's Inferno. A superplanckian inflationary trajectory can fit into a region of subplanckian diameter as shown in figure \ref{coord}. This would be impossible in a single field model\footnote{Of course invoking a non-canonically normalized single field does not help because the diameter of a region in field space has to be measured in terms of the field-space metric.}. Let us make this quantitative. In the $\cu$ direction the entire inflationary dynamics is contained in a region of diameter:
\be
d_{\cu}=\frac{\fu}{\sqrt{\fu^2+\fd^2}}\Delta \phi_{\rm eff}\; \simeq\; \frac{\fu}{ \fd}\Delta \phi_{\rm eff}\left[1-\mathcal{O}\left(\frac{\fu^2}{\fd^2}\right)\right]\,.
\ee
Thus, provided on can achieve a large enough hierarchy $\fu/\fd$, then $d_{\cu} < \Mpl \ll \Delta\phi_{\rm eff}$. The $\cd$ direction enjoys a shift symmetry, hence the largest diameter possible is the period itself, i.e.~$d_{\cd} = 2\pi \fd$. To summarize, $d_{\cu} \ll \Mpl$ and $d_{\cd}\ll\Mpl$ (from condition A). Hence we have shown that the entire high-scale chaotic-like inflationary dynamics can be contained in a region of subplanckian diameter. The discussions in this section applies \textit{mutatis mutandis} to any $p$. 

%%%%%%%%%%%%%%%%%%%%%%%%%%%%%%%%%%%%%%%%%%%%%%%%%%%%%%%%%

\section{The string theory model}

In this section we show how to embed Dante's Inferno in string theory. The construction is a simple generalization of the axion monodromy models constructed in \cite{Silverstein:2008sg,McAllister:2008hb,Flauger:2009ab}. We discuss only those ingredients that are new and special to the two-field model \eqref{V} and refer the reader to \cite{McAllister:2008hb,Flauger:2009ab} for further details on the explicit construction. The bottom line is that Dante's Inferno relaxes some constraints and dangerous backreaction effects present in single-field axion monodromy models. Roughly speaking, the importance of backreaction and other corrections turns out to be proportional to the range of variation of the axions involved in the model. Explicit examples are the backreaction of four-cycle volumes and light Kaluza-Klein masses discussed in \cite{Flauger:2009ab}. In Dante's Inferno, as opposed to the single-field case, the ranges of variation $d_{\cu}$ and $d_{\cd}$ of the axions, i.e. the fundamental fields in \eqref{V}, are small in Planck units. This holds even though the length of the inflationary trajecotry, as measured by the effective field $\pe$ in \eqref{p}, is superplanckian (inflation takes place at the GUT scale).

%%%%%%%%%%%%%%%%%%%%%%%%%%%%%%%%%%%%%%%%%%%%%%%%%%%%%%%%%%%%%

\subsection{Ingredients} 

The first step is to ensure that the two axionic fields in \eqref{V} are present in the low energy effective action of string theory. Following \cite{McAllister:2008hb,Flauger:2009ab}, we consider Type IIB Calabi-Yau 
orientifold compactifications. The integral of ten-dimensional two-form fields over two-cycles leads to four-dimensional axions which are in one-to-one correspondence with the homology classes $H_{(1,1)}^-$ of orientifold-odd two-cycles. Hence we require $\mathrm{dim} \, H_{(1,1)}^-\geqslant 2$. These axions enjoy a shift symmetry to all orders in perturbation theory that is broken only by either non-perturbative effects or by the presence of branes. 

We consider a situation in which Euclidean D1-brane (ED1) instantons are present with the right number of fermionic zero-modes (four) in order to correct the K\"ahler potential\footnote{At present, the precise details of these non-perturbative corrections are not known, but along the lines of \cite{Camara:2009xy}, we expect
there to be progress on this in the near future. It would be useful to know in more detail
in which cases they arise, and the precise conditions under which they give non-holomorphic contributions.
For the rather limited scope of this short note, the corrections merely need to exist and satisfy one requirement that we state below.}. Without loss of generality, we choose a basis $H_{(1,1)}^-$ such that the two-cycle $\Sigma_{\rm ED1}$ that supports the instanton is a linear combination of two basis elements $\Sigma_{\cu}$ and $\Sigma_{\cd}$ that correspond to the four-dimensional axions $\cu$ and $\cd$ respectively. This arrangement produces a cosine term as in \eqref{V} but with a possibly different argument:
\begin{equation}
\cos\left(\alpha \frac{\cu}{\fu}-\beta
\frac{\cd}{\fd }\right)
\end{equation}
for some real numbers $\alpha$ and $\beta$,
that are in principle determined by a calibration condition on $\Sigma_{\rm ED1}$. Now $\Lambda^4\propto e^{-S_{\rm ED1}}$ where $S_{\rm ED1}$ is the instanton action, which is proportional to the volume of $\Sigma_{\rm ED1}$. Higher instanton contributions are suppressed by further powers of $e^{-S_{\rm ED1}}$.

 \begin{figure}
\begin{center}
\includegraphics[width=60mm]{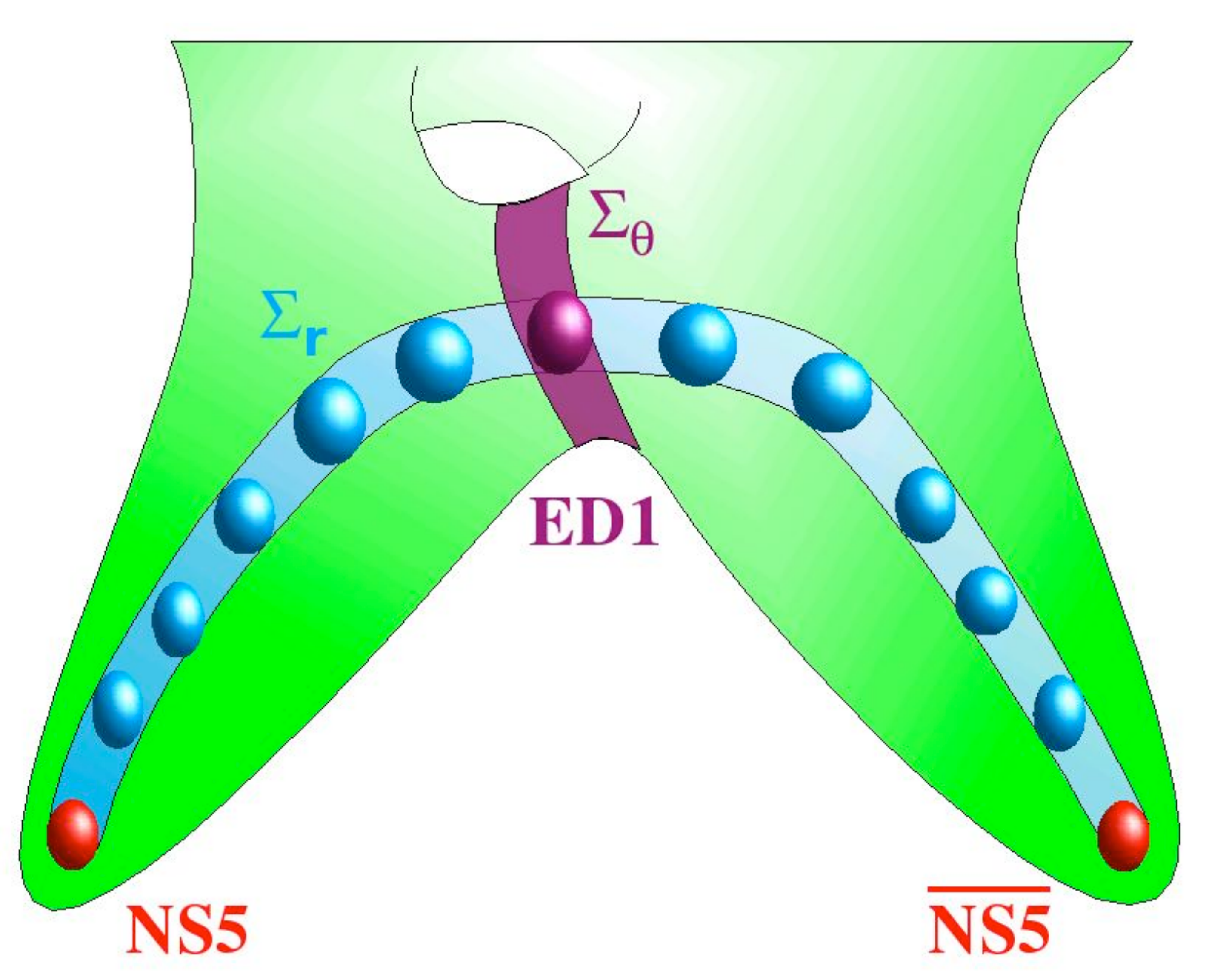}
\caption{A cartoon of the ingredients required to embed Dante's Inferno in string theory.}
\label{Mfig}
\end{center}
\end{figure}

The crucial assumption is that one can introduce a monodromy for one of the two axions, say $\cu$, in a controllable manner. A first attempt would be to assume that $\cu$ comes from the NSNS B-field and $\cd$ comes from the RR two-form field. In this case the shift symmetry of the B-type axion would be broken by the non-perturbative corrections to the superpotential required in a moduli stabilization \`a la KKLT \cite{Kachru:2003aw}, while the C-type axion would be unaffected. A non-perturbative effect generated by $(p,q)$-string instantons could depend on a linear combination of B- and C-type axions corresponding to the same cycle, with e.g.~$(\alpha,\beta)=(1,1)$ in the leading instanton contribution. But the axion decay constants for C-type axions turn out to be suppressed with respect to B-type axions by a factor of $g_s$. This makes it hard to ensure the hierarchy $\fu/\fd$ that was crucial for the mechanism described above, so this first attempt is unsuccessful.

An alternative that does work was proposed in \cite{McAllister:2008hb}: a monodromy for a C-type axion can be obtained wrapping an NS5-brane on the two-cycle that supports the axion. Another NS5-brane can wrap a homologous two-cycle with opposite orientation in order to cancel the tadpole. For example, the two NS5-branes could be located in two different warped throats as in figure \ref{Mfig}. Then a potential barrier prevents the classical instability, rendering the configuration metastable. As the height of the barrier is controlled by the warping, the quantum instability due to tunneling can be parametrically suppressed. In addition, warping provides
control of the supersymmetry breaking. Hence the crucial requirement is that the two-cycle 
$\Sigma_{\cu}$ is a member of a family of homologous two-cycles that extends into a warped throat as shown in figure \ref{Mfig}. Via S-duality we expect the induced monodromy term to be  
\be   \label{leq}
\frac{\epsilon}{g_s (2\pi)^5 \al}\sqrt{l^4+\left(\frac{2 \pi g_s \cu}{\fu}\right)^2} \simeq \mu^3 \cu + \mathrm{const}
\ee
where $\epsilon$ encodes the warping. In the single-field axion monodromy model it is important that (\ref{leq}) is the leading term in the potential. The analysis of \cite{Flauger:2009ab} shows that a potentially competing effect comes from the backreaction on the geometry of the D3-charge induced by the large vev of the inflaton. In the same paper, a resolution of this problem was proposed that requires both the NS5-brane and the tadpole-canceling anti-NS5-brane to be placed in a warped bifurcated throat such that the backreaction is due to a dipole as opposed to a monopole.
In the present model, we do not need to resort to such model-building tricks. 
The precise form of the monodromy term $W(\cu)$ is not crucial to the success of inflation, and in fact $W(\cu)$ could be a steep potential that by itself would not give rise to slow-roll dynamics. Had we overlooked an effect that would induce an $\eta$ problem in the $\cu$ direction, we would simply need to consider a larger hierarchy $\fu/\fd$. The Dante's Inferno setup hence alleviates the difficulty of precisely determining the leading breaking of the shift symmetry.

Let us now turn to the question of genereting the small hierarchy $\fu/\fd$ in the first place. In the string theory construction, the axion decay constant for a C-type axion corresponding to a generic two-cycle $\Sigma$ is \cite{Grimm:2004uq,Flauger:2009ab}
\be 
\frac{f^2_{\Sigma}}{\Mpl^2}=\frac{g_s}{8 \pi^2} \frac{c_{\alpha \Sigma\Sigma} v^{\alpha}}{\V}\,,
\ee
where $c_{\alpha \beta \gamma}$ are intersection numbers, $\V$ is the volume of the compact space and $v^\alpha$ are two-cycle volumes. Hence, the condition that needs to hold is
\be 
\frac{\fu}{\fd}=\frac{\beta}{\alpha}\frac{c_{\gamma r r} v^{\gamma}}{c_{\gamma \theta \theta} v^{\gamma}}\ll 1\,, \label{con}
\ee
where again $\alpha$ and $\beta$ are the coefficients of the linear combination of two-cycles involved in the non-perturbative effect. As stated earlier, in a typical situation a ratio $\fu/\fd\sim\mathcal{O}(10^{-1}-10^{-2})$ is enough to give 60 e-folds of slow-roll inflation within a region of subplanckian diameter. Without any further information, 
one could guess that roughly half of the time $\alpha>\beta$, which helps our mechanism work. Even if $\alpha<\beta$, it
is only when $\beta/\alpha$ is a large number that our mechanism would fail. In typical compactifications, intersection numbers and two-cycle volumes can both easily range over one or two orders of magnitude. Hence, we believe that \eqref{con} can be easily satisfied. Further work on ED1 instantons should help clarify this situation. 

As a side remark, there is another possibility to construct a small ratio $\fu/\fd$. Recall that $\cu$ corresponds to a two-cycle that must extend into a warped region (otherwise the NS5-brane would break supersymmetry at the string scale). Then $\fu$ is suppressed by a factor $e^{-A_{top}}$ \cite{McAllister:2008hb}, which is the largest value taken by the warp factor in the family of two-cycles homologous to $\Sigma_r$. This warping suppression could account for the required hierarchy in cases where the ingredients in \eqref{con} do not offer enough flexibility.

One can always contend that the small ratio  $\fu/\fd$ that we 
used to suppress corrections may itself be unstable to corrections. 
This was not investigated in \cite{McAllister:2008hb,Flauger:2009ab},
since there was only one $f$. 
We believe that the case for this to be stable is better than, for example, the case for the inflaton mass to be stable. 
It would be interesting to investigate this in more detail, however, for example
in toy string models. 

To summarize, we have shown how to embed Dante's Inferno in string theory as a straightforward generalization of the axion monodromy model in \cite{McAllister:2008hb,Flauger:2009ab}. The main difference compared to earlier work 
 is that we consider the situation where non-perturbative effects (such as ED1 instantons) and monodromy-inducing effects (such as wrapped NS5-branes) 
 are associated with two-cycles that partially overlap  but are globally distinct. As far as we can see, there is no  obstruction in principle to the existence of 
this class of configurations. We leave the study of more detailed examples for future work. 

%%%%%%%%%%%%%%%%%%%%%%%%%%%%%%%%%%%%%%%%%%%%%%%%%%%%%%%%%%%%%

\subsection{Constraints}

The main computability constraints of our model are: 
\begin{itemize}
\item $g_s\ll 1$ for string theory to be perturbative.
\item $V\ll\U$, where $\U$ is a moduli-stabilization barrier, for the inflationary dynamics not to destroy moduli stabilization.
\item Volumes of cycles large in string units, $v \gg 1$, in order to neglect world-sheet instantons and $\al$ corrections.
\item Small (but not negligible) non-perturbative corrections, in order to neglect higher instanton
contributions.
\end{itemize}
We have checked that these constraints are satisfied in the same way as they are in the class of single field axion monodromy models studied in \cite{McAllister:2008hb,Flauger:2009ab}. 

More constraints come from the consistency of the string theory construction. The key observation here is that these constraints arise when an axion with a mono\-dromy receives a large vev. In Dante's Inferno this vev is parametrically smaller  than in single-field axion mono\-dromy, by a factor $\fu/\fd$. Hence each constraint is relaxed by factors of this form. 

First we consider the backreaction on the geometry, mainly due to the large induced D3-brane charge on the NS5-brane.  We notice that the induced effective D3-brane charge is 
\be 
N_{D3}=\frac{\cu}{2\pi \fu}\,.
\ee
This induced charge is now parametrically smaller than in the single-field axion monodromy case, because the $\cu$ range required for 60 e-folds of inflation is now
(taking the linear case for definiteness) $11\Mpl f_r/f_{\theta}$ as opposed to $11\Mpl$. This extends the possible parameter range of the model as compared to the discussion in \cite{Flauger:2009ab}. 

A second concern is the backreaction on the warp factor that in turn changes the four-cycle volumes. The latter are stabilized, and hence this interplay can generate an $\eta$ problem for single-field axion monodromy. Two different model building tricks to alleviate this problem were proposed in \cite{Flauger:2009ab}. For the present two-field model this issue can conceivably be solved without resort to such tricks. To see this, recall that the inflaton direction is not $\cu$ but a linear combination of $\cu$ and $\cd$. Even if the $\cu$ direction sees
its approximate flatness destroyed by corrections, it is likely that slow-roll inflation still proceeds because the mass for the effective inflaton is suppressed with respect to the mass of $\cu$ by a small factor of the order $\fu/\fd$. This is the reason why in the previous sections we stressed that our mechanism works for quite general monodromy terms $W(r)$ in \eqref{V}.

Finally, a last concern comes from the KK modes on the world-volume of the NS5-brane that induces the monodromy for $\cu$. These become light due to the large flux and an estimate \cite{Flauger:2009ab} gives 
\be 
m^2_{{\rm KK, NS5}}=\frac{g_2}{g_2+(\cu/\fu)^2}m_{\rm KK}^2,
\ee
where $g_2$ is the determinant of the metric on the two-cycle wrapped by the NS5-brane. This effect can be intuitively understood in a T-dual picture where a brane with flux becomes a tilted brane. The larger the flux, the more times the T-dual brane winds around, and
the  longer it becomes. The KK modes in this long direction become increasingly light. Again, in Dante's Inferno, $\cu$ takes parametrically smaller vacuum expectation values than in the single-field case, which ensures heavier KK masses.

%%%%%%%%%%%%%%%%%%%%%%%%%%%%%%%%%%%%%%%%%%%%%%%%%%%%%%%%%

\section*{Acknowledgments}

We would like to thank Ulf Danielsson, Michael Haack, Liam McAllister, Henry Tye, Bret Underwood and Gang Xu for helpful comments. The research of E.P. was supported in part by the National Science Foundation through grant NSF-PHY-0757868. M.B. thanks the Swedish Research Council (VR) for support. We are also grateful to the Swedish Foundation for International Cooperation in Research and Higher Education (STINT) for partial support.

%%%%%%%%%%%%%%%%%%%%%%%%%%%%%%%%%%%%%%%%%%%%%%%%%%%%%%%%%


\begin{thebibliography}{99}

\bibitem{Lyth:1996im}
  D.~H.~Lyth,
  ``What would we learn by detecting a gravitational wave signal in the  cosmic
  microwave background anisotropy?,''
  Phys.\ Rev.\ Lett.\  {\bf 78} (1997) 1861
  [arXiv:hep-ph/9606387].
  %%CITATION = PRLTA,78,1861;%%

\bibitem{Dante}
D.~Alighieri, ``Commedia" (1304-1321).
We recommend
G.\ Petrocchi, ed., ``La Commedia secondo l'antica vulgata". Milan, Italy: Mondadori (1966). 

%\cite{Kachru:2002gs}
\bibitem{Banks:2003sx}
  T.~Banks, M.~Dine, P.~J.~Fox and E.~Gorbatov,
  ``On the possibility of large axion decay constants,''
  JCAP {\bf 0306}, 001 (2003)
  [arXiv:hep-th/0303252].
  %%CITATION = JCAPA,0306,001;%%
  
  
%   \bibitem{Baumann:2009ds}
%   D.~Baumann,
%   ``TASI Lectures on Inflation,''
%   arXiv:0907.5424 [hep-th].
%   %%CITATION = ARXIV:0907.5424;%%

%\cite{Grimm:2004uq}
\bibitem{Grimm:2004uq}
  T.~W.~Grimm and J.~Louis,
  ``The effective action of N = 1 Calabi-Yau orientifolds,''
  Nucl.\ Phys.\  B {\bf 699}, 387 (2004)
  [arXiv:hep-th/0403067].
  %%CITATION = NUPHA,B699,387;%%

%\cite{Baumann:2009ni}
\bibitem{Kim:2004rp}
  J.~E.~Kim, H.~P.~Nilles and M.~Peloso,
  ``Completing natural inflation,''
  JCAP {\bf 0501} (2005) 005
  [arXiv:hep-ph/0409138].
  %%CITATION = JCAPA,0501,005;%%

\bibitem{Freese:1990rb}
  K.~Freese, J.~A.~Frieman and A.~V.~Olinto,
  ``Natural inflation with pseudo - Nambu-Goldstone bosons,''
  Phys.\ Rev.\ Lett.\  {\bf 65} (1990) 3233.
  %%CITATION = PRLTA,65,3233;%%


\bibitem{Silverstein:2008sg}
  E.~Silverstein and A.~Westphal,
  ``Monodromy in the CMB: Gravity Waves and String Inflation,''
  Phys.\ Rev.\  D {\bf 78} (2008) 106003
  [arXiv:0803.3085 [hep-th]].
  %%CITATION = PHRVA,D78,106003;%%

\bibitem{McAllister:2008hb}
  L.~McAllister, E.~Silverstein and A.~Westphal,
  ``Gravity Waves and Linear Inflation from Axion Monodromy,''
  arXiv:0808.0706 [hep-th].
  %%CITATION = ARXIV:0808.0706;%%

 \bibitem{Flauger:2009ab}
  R.~Flauger, L.~McAllister, E.~Pajer, A.~Westphal and G.~Xu,
  ``Oscillations in the CMB from Axion Monodromy Inflation,''
  arXiv:0907.2916 [hep-th].


\bibitem{Camara:2009xy}
%\cite{Camara:2008zk}
%\bibitem{Camara:2008zk}
 P.~G.~Camara and E.~Dudas,
 ``Multi-instanton and string loop corrections in toroidal orbifold models,''
 JHEP {\bf 0808} (2008) 069
 [arXiv:0806.3102 [hep-th]].
 %%CITATION = JHEPA,0808,069;%%
P.\ Camara, E. Dudas, work in progress. 
  
\bibitem{Kachru:2003aw}
  S.~Kachru, R.~Kallosh, A.~D.~Linde and S.~P.~Trivedi,
  ``De Sitter vacua in string theory,''
  Phys.\ Rev.\  D {\bf 68}, 046005 (2003)
  [arXiv:hep-th/0301240].
 

\end{thebibliography}
\end{document}